\documentclass[apjl]{emulateapj}

\usepackage{graphicx}
\usepackage{graphics}

\newcommand{\name}{ASASSN-13db}
\newcommand{\starname}{SDSSJ0510}
\newcommand{\swift}{{\it Swift}}
\newcommand{\msun}{\ensuremath{\rm{M}_\odot}}
\newcommand{\Lsun}{\ensuremath{\rm{L}_\odot}}
\newcommand{\rsun}{\ensuremath{\rm{R}_\odot}}

\begin{document}
\title{Discovery and Observations of ASASSN-13db, an EX~Lupi-Type Accretion Event on a Low-Mass T Tauri Star}
\shorttitle{ASASSN-13db}

\author{Thomas~W.-S.~Holoien\altaffilmark{*,1}, {Jose~L.~Prieto}\altaffilmark{2,3}, {K.~Z.~Stanek}\altaffilmark{1,4},
{C.~S.~Kochanek}\altaffilmark{1,4},
{B.~J.~Shappee}\altaffilmark{1},
{Z.~Zhu}\altaffilmark{2,5},
{A.~Sicilia-Aguilar}\altaffilmark{6,7},
{D.~Grupe}\altaffilmark{8},
{K.~Croxall}\altaffilmark{1},
{J.~J.~Adams}\altaffilmark{9},
{J.~D.~Simon}\altaffilmark{9},
{N.~Morrell}\altaffilmark{10},
{S.~M.~McGraw}\altaffilmark{11},
{R.~M.~Wagner}\altaffilmark{1,12},
{U.~Basu}\altaffilmark{1,13},
{J.~F.~Beacom}\altaffilmark{1,4,14},
{D.~Bersier}\altaffilmark{15},
{J.~Brimacombe}\altaffilmark{16},
{J.~Jencson}\altaffilmark{1},
{G.~Pojmanski}\altaffilmark{17},
{S.~G.~Starrfield}\altaffilmark{18},
{D.~M.~Szczygie{\l}}\altaffilmark{17},
{C.~E.~Woodward}\altaffilmark{19}
}

\altaffiltext{*} {tholoien@astronomy.ohio-state.edu}
\altaffiltext{1}{Department of Astronomy, The Ohio State University, 140 West 18th Avenue, Columbus, OH 43210, USA}
\altaffiltext{2}{Department of Astrophysical Sciences, Princeton University, 4 Ivy Lane, Peyton Hall, Princeton, NJ 08544, USA}
\altaffiltext{3}{Carnegie-Princeton Fellow}
\altaffiltext{4}{Center for Cosmology and AstroParticle Physics (CCAPP), The Ohio State University, 191 W.\ Woodruff Ave., Columbus, OH 43210, USA}
\altaffiltext{5}{Hubble Fellow}
\altaffiltext{6}{Departamento de F\'{i}sica Te\'{o}rica, Facultad de Ciencias, Universidad Aut\'onoma de Madrid, 28049 Cantoblanco, Madrid, Spain}
\altaffiltext{7}{SUPA, School of Physics and Astronomy, University of St. Andrews, North Haugh, St. Andrews KY16 9SS, Scotland, UK}
\altaffiltext{8}{Department of Astronomy and Astrophysics, Pennsylvania State University, 525 Davey Lab, University Park, PA 16802, USA}
\altaffiltext{9}{Observatories of the Carnegie Institution for Science, 813 Santa Barbara St., Pasadena, CA 91101, USA}
\altaffiltext{10}{Carnegie Observatories, Las Campanas Observatory, Colina El Pino, Casilla 601, Chile}
\altaffiltext{11}{Department of Physics and Astronomy, Ohio University, 251B Clippinger Labs, Athens, OH 45701, USA}
\altaffiltext{12}{Large Binocular Telescope Observatory, University of Arizona, 933 N Cherry Avenue, Tucson, AZ 85721, USA}
\altaffiltext{13}{Grove City High School, 4665 Hoover Road, Grove City, OH 43123, USA}
\altaffiltext{14}{Department of Physics, The Ohio State University, 191 W. Woodruff Ave., Columbus, OH 43210, USA}
\altaffiltext{15}{Astrophysics Research Institute, Liverpool John Moores University, 146 Brownlow Hill, Liverpool L3 5RF, UK}
\altaffiltext{16}{Coral Towers Observatory, Cairns, Queensland 4870, Australia}
\altaffiltext{17}{Warsaw University Astronomical Observatory, Al. Ujazdowskie 4, 00-478 Warsaw, Poland}
\altaffiltext{18}{School of Earth and Space Exploration, Arizona State University, Box 871404, Tempe, AZ 85287-1404, USA}
\altaffiltext{19}{Minnesota Institute for Astrophysics, University of Minnesota, 116 Church St., SE, Minneapolis, MN 55455, USA}

\begin{abstract}
We discuss ASASSN-13db, an EX~Lupi-type (``EXor'') accretion event on the young stellar object (YSO) SDSS J051011.01$-$032826.2 (hereafter {\starname}) discovered by the All-Sky Automated Survey for SuperNovae (ASAS-SN). Using archival photometric data of {\starname} we construct a pre-outburst spectral energy distribution (SED) and find that it is consistent with a low-mass class II YSO near the Orion star forming region ($d \sim 420$~pc). We present follow-up photometric and spectroscopic observations of the source after the $\Delta V \sim-$5.4 magnitude outburst that began in September 2013 and ended in early 2014. These data indicate an increase in temperature and luminosity consistent with an accretion rate of $\sim10^{-7}$~${\msun}$~yr$^{-1}$, three or more orders of magnitude greater than in quiescence. Spectroscopic observations show a forest of narrow emission lines dominated by neutral metallic lines from \ion{Fe}{1} and some low-ionization lines. The properties of ASASSN-13db are similar to those of the EXor prototype EX~Lupi  during its strongest observed outburst in late 2008. 
\end{abstract}

\keywords{stars: activity --- stars: individual (SDSS J051011.01$-$032826.2) --- stars: pre-main sequence}

\section{Introduction}
\label{sec:intro}

Young stars are often variable \citep[e.g.,][]{herbst94}, and are known to undergo significant eruptions marking substantial increases in their accretion rates \citep[e.g.,][]{herbig77, hartmann96}, during which they are thought to accrete a significant fraction of their total mass. A class of young stellar objects (YSOs) that repeatedly undergo these eruptions, named ``EXors'' after the prototype EX~Lupi \citep[e.g.,][]{herbig89, herbig08}, has been the subject of recent studies \citep[e.g.,][]{Sicilia2012, lorenzetti12, antoniucci2013, kospal14}. The physical mechanisms responsible for triggering these outbursts remain poorly understood \citep[e.g.,][]{Sicilia2012, kospal14}, but a recent study of the 2008 EX~Lupi outburst by \citet{kospal14} suggests that a sub-stellar binary companion could be involved. Other possible causes include the magnetorotational instability \citep[e.g.,][]{bae13}, disk fragmentation \citep[e.g.,][]{vorobyov05}, and instabilities at the disk truncation radius from the stellar magnetosphere \citep[e.g.,][]{dangelo10}.

Among the defining characteristics of EXors are that they erupt every few years with variable peak luminosities, typical visual amplitudes of up to $\sim 4$~mag, and outburst durations of a few months to a year. At minimum light, EXors resemble classical T Tauri stars (TTSs) with absorption spectra like those of late K- or early M-dwarfs and a strong \ion{Li}{1} line at 6707\AA~\citep{herbig08}. At maximum light, their spectra are dominated by a hot continuum and quiescent absorption features appear as emission lines, indicating increased accretion \citep{herbig89, herbig08, lorenzetti12}.

In this Letter we describe the discovery and follow-up observations of a new EXor-type outburst, {\name}. The transient was discovered by the All-Sky Automated Survey for SuperNovae (ASAS-SN\footnote{\url{http://www.astronomy.ohio-state.edu/$\scriptstyle\mathtt{\sim}$assassin/index.shtml}}), a long-term project to monitor the whole sky on a rapid cadence to find nearby supernovae and other transients (see \citealt{shappee13} for details). Our transient source detection pipeline was triggered on 2013 September 24, detecting a change in $V$-band magnitude from $V\sim19.4$ to $V\sim14$ over a span of $\sim$~15 days in the red source SDSS J051011.01$-$032826.2 (hereafter {\starname}). The source is located roughly $6.6^{\circ}$ ($\sim$ 50 pc) from the Orion Nebula and roughly $1.4^{\circ}$ ($\sim$~10 pc) from the L1615/L1616 cloud, both regions of star formation at a distance of $\sim420$ pc (averaging recent estimates from \citet{menten07, hirota07}; and \citet {sandstrom07}). We assume this distance for {\starname}. Follow-up spectroscopic observations with the Magellan Echellette spectrograph (MagE; \citealt{Marshall2008}) on 2013 September 26 showed a highly unusual spectrum with a plethora of emission lines, and we began a follow-up campaign in order to characterize this transient.

In \S\ref{sec:obs} we describe pre-outburst archival observations, including a previous outburst recorded in 2010$-$2011, and new data taken during and after the outburst as part of our follow-up campaign. In \S\ref{sec:analysis} we analyze these data and describe the properties of the star in quiescence and in outburst. Finally, in \S\ref{sec:disc} we compare these properties to those of known EXors and the prototype EX~Lupi.

\section{Observations and Survey Data} 
\label{sec:obs} 

Here we summarize the available archival data in quiescence and outburst as well as our new photometric and spectroscopic observations.

\subsection{Photometry in Quiescence}
\label{sec:quiescence}
We retrieved archival photometry of {\starname} in quiescence spanning the optical to the mid-infrared (IR) from the Sloan Digital Sky Survey \citep[SDSS;][]{york00}, the Two-Micron All Sky Survey \citep[2MASS;][]{skrutskie06}, and the Wide-field Infrared Survey Explorer \citep[WISE;][]{wright10} based on a coordinate cross-match to the ASAS-SN position. There are no Spitzer, Herschel, or AKARI observations of the source. We used the $ugriz$, $JHK_S$, and $W1W2W3W4$ magnitudes in \S\ref{sec:sed} to construct a Spectral Energy Distribution (SED) for the progenitor, as shown in Figure~\ref{fig:sed}. From the SDSS data, and using M-dwarf templates from \citet{bochanski07}, we estimate the quiescent V-band magnitude to be $V\sim19.4$.

\subsection{Prior Outburst}
\label{sec:prioroutbursts}
After the discovery of {\name}, we examined existing data for {\starname} from the Catalina Real-Time Transient Survey \citep[CRTS;][]{drake09}, the Mobile Astronomical System of TElescope Robots (MASTER), and the Near-Earth Asteroid Tracking program (NEAT). The NEAT data indicate that {\starname} did not have any eruptions between December 2000 and February 2003, but we found that it had a previously unreported outburst in late 2010 and early 2011 that was observed by CRTS and MASTER (Denisenko et al., private communication). {\starname} had $V>17$ on 2010 November 08 (CRTS) and 2011 October 2 (CRTS), but had $V\sim15.1$ in 2010 December (multiple CRTS epochs) and was brighter than the present outburst with $V\sim13.2$ on 2011 January 22 (MASTER). The length and magnitude of this previous outburst are consistent with a large EXor-type event.

\subsection{New Photometric Observations}
\label{sec:phot}

The emission lines seen in the initial follow-up spectrum of {\name} indicated that it must be hot and should have increased ultraviolet (UV) emission. Based on this, we requested and were granted a series of {\swift} X-ray Telescope \citep[XRT;][]{burrows05} and UltraViolet and Optical Telescope \citep[UVOT;][]{roming05} ToO observations. The {\swift} UVOT observations of {\name} were obtained in 6 filters: $V$ (5468~\AA), $B$ (4392~\AA), $U$ (3465~\AA), $UVW1$ (2600~\AA), $UVM2$ (2246~\AA), and $UVW2$ (1928~\AA) \citep{poole08}. We used the UVOT software task  {\it uvotsource} to extract the source and background counts from a 5\farcs0 radius region and a sky region with radius of $\sim$20\farcs0. The UVOT count rates were converted into magnitudes and fluxes based on the most recent UVOT calibration \citep{poole08, breeveld10}. The UVOT data are shown along with other photometric data in Figure~\ref{fig:lightcurve}.

The XRT was operating in Photon Counting mode \citep{hill04} during our observations. The data were reduced and combined with the software tasks {\it xrtpipeline} and {\it xselect} to obtain an image in the 0.3$-$10 keV range with total exposure time of 5388 s. We used a region with a radius of 20 pixels (47\farcs{1}) centered on the source position to extract source counts and a source-free region with a radius of 100 pixels (235\farcs{7}) for background counts. We measured 7 counts in the source region and 113 counts in the background region, giving an expected number of counts in the source region of 4.5. We do not detect X-ray emission from {\name} to a 3-sigma upper limit \citep{kraft91} of $2.5\times10^{-3}$~counts s$^{-1}$, equivalent to $\sim 1.3\times10^{-13}$~erg cm$^{-2}$ s$^{-1}$. This is not strongly constraining, as the source's emission drops strongly at sub-UV wavelengths, so we do not use this limit when fitting the outburst SED.

In addition to the {\swift} observations, we obtained optical and near-IR photometry from a number of ground-based observatories. $VRI$ observations were obtained at the Las Cumbres Observatory Global Telescope Network (LCOGT) 1-m facilities in Cerro Tololo, Chile and
Sutherland, South Africa \citep{brown13} and with the Ohio State Multiobject Spectrograph (OSMOS; \citealt{Martini2011}) on the MDM 2.4-m telescope. $VRIJHK$ observations were obtained with ANDICAM on the CTIO 1.3-m telescope. All images were reduced following standard
procedures. The $VRI$ aperture magnitudes were calibrated using SDSS stars in the field transformed onto the Johnson-Cousins magnitude system using \citet{lupton05}, while the $JHK$ magnitudes were calibrated using a 2MASS star in the field. These data are shown in Figure~\ref{fig:lightcurve}, along with all photometric data from the ASAS-SN pipeline.

\subsection{New Spectroscopic Observations}
\label{sec:spec}

We obtained optical spectra with multiple telescopes and instruments between 2013 September 26 and 2014 January 19. The instruments/telescopes used were CCDS on the MDM~2.4-m, the Magellan Echellette spectrograph (MagE) on the Magellan Clay 6.5-m at Las Campanas Observatory (LCO), the Multi-Object Double Spectrographs (MODS; \citealt{Pogge2010}) on the 8.4-m Large Binocular Telescope (LBT) on Mount Graham, the Dual Imaging Spectrograph (DIS) on the Apache Point Observatory 3.5-m telescope, and the B\&C spectrograph on the Steward
Observatory Bok 2.3-m telescope at the Kitt Peak National Observatory (KPNO). The spectra from CCDS, DIS, and B\&C were reduced using standard techniques in IRAF. The spectra from MODS were reduced using IRAF and the MODS pipeline in IDL. The spectra from MagE were reduced using the Carnegie pipeline. Figure~\ref{fig:specevol} shows a montage of the flux-calibrated spectra from MODS, CCDS, and the B\&C. Figure~\ref{fig:speccomp} shows a small section of the MagE intermediate-resolution ($\rm R\sim 5000$) spectra. 

\section{Analysis}
\label{sec:analysis}

\subsection{Spectral Energy Distribution}
\label{sec:sed}

\begin{figure}[]
\centering
\includegraphics[width=0.8\linewidth]{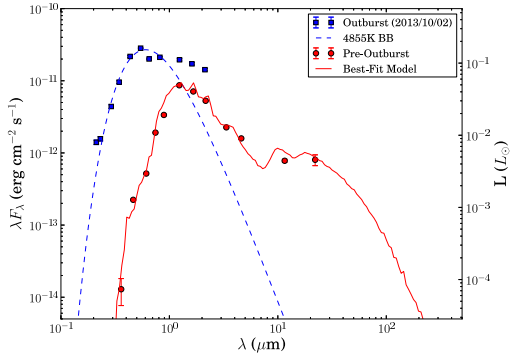} 
\caption{Spectral energy distribution of {\starname} before and during its outburst. The red circles indicate pre-outburst photometric data and the blue squares indicate outburst data. Error bars are shown, but in most cases they are smaller than the points. The solid red curve shows the best fit to the pre-outburst data from the models of \citet{robitaille07} (model ID 3016557). We also estimate an additional thermal emission component by fitting a $T=4855$ K blackbody curve to the outburst data (dashed blue curve).}
\label{fig:sed}
\end{figure}

Figure~\ref{fig:sed} shows the pre-outburst SED of {\starname} as compared to its SED during outburst on 2013 October 2. Also shown is the best-fit SED model for {\starname} generated by the \citet{robitaille07} YSO SED fitting tool\footnote{\url{http://caravan.astro.wisc.edu/protostars/sedfitter.php}} (model ID 3016557). The fitting method compares the observed SED to 200,000 model YSO SEDs consisting of pre-main sequence stars with various combinations of stellar parameters embedded in circumstellar disks with outflow cavities in order to sample a large range of parameter space. To account for any variability of {\starname} in quiescence, we used 10\% minimum uncertainties on all input magnitudes. 

\begin{deluxetable*}{lccc} \tablewidth{0pt} \tabletypesize{\scriptsize}
\tablecaption{Properties of {\starname} Before and During Outburst}
\tablehead{ \colhead{ } & \colhead{Best-Fit Model (Pre-Outburst)} & \colhead{$\Delta \chi^2 < 3$ Range (Pre-Outburst)} & \colhead{During Outburst}}
\startdata
Stellar Age (yr)  &  $3.5\times10^6$ & $1\times10^6-2\times10^7$ & --- \\
Stellar Mass ({\msun})  &  0.15  & $0.1-0.4$ &  --- \\
Stellar Radius ({\rsun})  &  1.1  &  $0.7-1.1$ & --- \\
Effective Temperature (K)  &  3075  & $2700-3900$ & 4855 \\
Total Luminosity ({\Lsun})  &  0.1  &$0.04-0.4$ & $>0.2$ \\
Accretion Luminosity ({\Lsun}) & $<4\times10^{-4}$ & --- & 0.2 \\
Disk Accretion Rate (${\msun}$~yr$^{-1}$) & $2\times10^{-12}$ & $10^{-12}-10^{-9}$ & $\sim10^{-7}$ \\
Disk Mass (${\msun}$) & $5.6\times10^{-5}$ & $ 4\times10^{-7}-6\times10^{-3}$ & --- \\
Disk Inner Radius (AU)  &  0.6  & $0.01-3$ &  --- \\
Disk Outer Radius (AU)  &  900  & $9-8000$ &  --- \\ 
Disk Scale Height at 100 AU (AU)  & 7 & $1-12$ & --- \\
Disk Flaring Power & 1.1 & $0.9-1.2$ & --- \\
\enddata
\label{table:props}
\end{deluxetable*}

The properties of {\starname} before and during outburst are summarized in Table~\ref{table:props}. The pre-outburst SED is well-modeled as a class II YSO at a distance of $\sim420$ pc. The best-fit model is consistent with a luminosity of $\sim0.1$~${\Lsun}$, $\rm T_{eff}\sim3100$~K, and $\rm M\star \sim 0.15$~M$_{\odot}$.  In addition to the properties of the best-fit model, we also list the minimum and maximum values for each parameter taken from all models with $\chi^2-\chi_{min}^2<3$ for all individual data points, as suggested in \citet{robitaille07}. While most parameters are well-constrained, some are not. For example, the best fitting accretion rate is $\dot{M}=2\times10^{-12}$ ${\msun}$~yr$^{-1}$, but realistically this should be viewed as an upper limit of $\dot{M}\lesssim10^{-10}$ ${\msun}$~yr$^{-1}$. Similarly, the distribution of inner disk radii is bimodal, at 0.02 and 0.5 AU, and only the latter is consistent with estimates for EX~Lupi \citep{sipos09}. None of the models are consistent with {\starname} having an envelope.

Other EXors in outburst have shown additional thermal emission \citep{lorenzetti12,Sicilia2012}. Fitting a blackbody curve to the outburst SED assuming the same extinction as before the outburst ($A_{V}=0.04$ for the best-fit model), we find that the bluer outburst data are well-fit by a
$T=4855$~K blackbody curve, slightly higher than the typical range of $1000$~K$-4500$~K \citep{lorenzetti12}. The near-IR outburst
data are not well-fit by this blackbody curve, indicating the presence of increased disk emission, but we lack the mid-to-far-IR data needed to fit an additional disk component. Integrating over the blackbody curve gives a lower limit to the total luminosity in outburst of $L=0.2$~${\Lsun}$.

Assuming the total luminosity in outburst is equivalent to the accretion luminosity, we estimate the mass accretion rate of {\name} in outburst using two methods. First, we use the scaling from \citet{natta04} based on the width of the H$\alpha$ line at 10\% of maximum. We measure this width to be $\sim$ 600 km~s$^{-1}$, which gives an accretion rate of $\dot{M}=9\times10^{-8}$~${\msun}$~yr$^{-1}$. We also use our estimate of the accretion luminosity combined with the stellar mass, stellar radius, and inner disk radius estimates from the SED fits and the method described in \citet{dahm08}. This gives a lower limit of $\dot{M}=5.5\times10^{-8}$~${\msun}$~yr$^{-1}$. These two estimates are mutually consistent, and we conclude that $\dot{M}\sim 10^{-7}$~${\msun}$~yr$^{-1}$ is a reasonable estimate of the accretion rate during the outburst.

The SED of the source in quiescence and during outburst are both consistent with {\name} being a large EXor accretion event on a low-mass TTS \citep[e.g.,][]{dalessio99,sipos09}, similar to the 2008 outburst of EX~Lupi \citep[e.g.,][]{herbig08,lorenzetti12}.

\subsection{Light Curve of {\name}}
\label{sec:lc}

\begin{figure}[]
\centering
\includegraphics[width=0.8\linewidth]{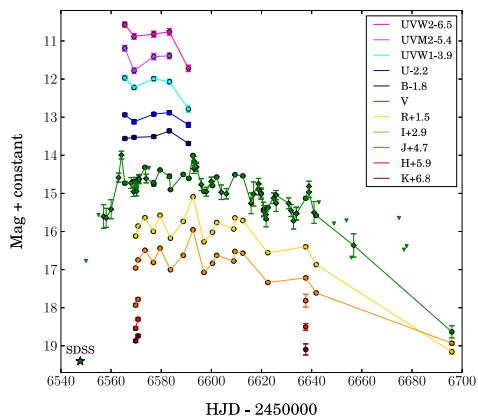} 
\caption{Light curve of {\name}, starting two weeks prior to discovery and spanning $\sim$ 4.5 months. Data obtained from {\swift} (UV $+$ optical), LCOGT 1-m (optical), MDM 2.4-m (optical), and SMARTS 1.3-m (optical $+$ near-IR) are shown as circles while data from ASAS-SN ($V$-band) are shown as diamonds. 3-sigma upper limits are shown as triangles for epochs where the source was not detected. The approximate quiescent V-band magnitude calculated from SDSS data is shown as a star in the bottom-left. The source exhibited day-to-day fluctuations of up to a magnitude and remained $4-5.5$ magnitudes brighter in $V$-band than in quiescence for the duration of the outburst. A table containing all photometric data is available in machine-readable form in the online journal.}
\label{fig:lightcurve}
\end{figure}

Figure~\ref{fig:lightcurve} shows the UV, optical, and near-IR light curves of {\name} from HJD 2546550 ($\sim1$ week prior to detection) through our latest epoch of observations on HJD 2546695. The ASAS-SN data show {\starname} brightening by $\Delta V\sim3.8$ magnitudes, at which point it was flagged by the ASAS-SN detection pipeline. Subsequent follow-up observations show the transient reaching a peak magnitude of $V\simeq14$ and fluctuating by up to $\sim1$ magnitude across all observed bands on timescales as short as $<1$ day. These variation timescales and amplitudes are consistent with other EXor events \citep[e.g.,][]{lorenzetti12}. {\starname} showed little variation in NIR colors before and during the outburst, implying the apparent brightening is likely not due a change in circumstellar extinction. During the $\sim4$-month outburst, {\starname} was consistently $\sim4-5.5$ magnitudes brighter than in quiescence. The star began to exhibit lower average magnitudes beginning around HJD 2456622, and has faded to $V\sim18.6$ on HJD 2456695, indicating the outburst has likely ended. 

\subsection{Spectral Analysis of {\name}}
\label{sec:specanal}

\begin{figure}[]
\centering
\includegraphics[width=0.8\linewidth]{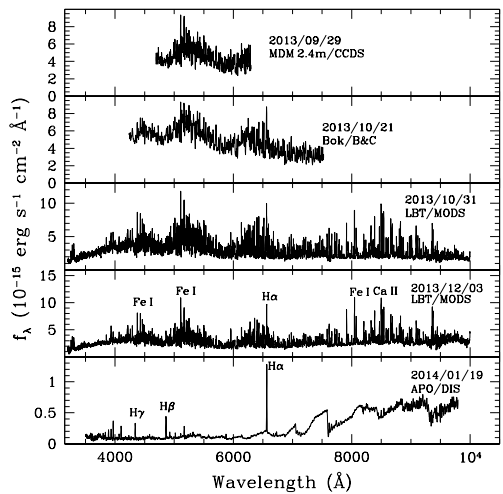}
\caption{Spectral time-sequence of {\name} during and after the outburst. Each spectrum shows the UT date and telescope/instrument used. During the outburst, the spectra are dominated by a forest of emission lines, mostly neutral metallic lines from \ion{Fe}{1}, with later epochs showing stronger features. Some of the strongest features, such as  \ion{Fe}{1}, H$\alpha$, and the \ion{Ca}{2} infrared triplet, are identified in the 2013 December 3 MODS spectrum. In contrast, the 2014 January 19 spectrum shows features characteristic of a TTS of M5 spectral type, indicating the outburst has likely ended. Note the change in the flux scale between panels.}

\label{fig:specevol}
\end{figure}

The spectral time-sequence of ASASSN-13db is shown in Figure~\ref{fig:specevol}. The spectra are completely dominated by a forest of emission lines during the outburst. The lines are mostly neutral metallic lines from \ion{Fe}{1}, with some low-ionization lines, including \ion{Fe}{2}, \ion{Ca}{2}, and \ion{Ti}{2}. Figure~\ref{fig:specevol} shows some of the strongest lines identified in the MODS spectrum from 2013 October 31, which include \ion{Fe}{1}, H$\alpha$, and the \ion{Ca}{2} infrared triplet. In contrast, the latest spectrum shows a strong red continuum and features characteristic of a TTS (M5 spectral type), again indicating the outburst has likely ended. In this spectrum we detect the $6707$~\AA~\ion{Li}{1} line, but we are unsure whether {\starname} is in a fully quiescent state.

\begin{figure}[]
\plottwo{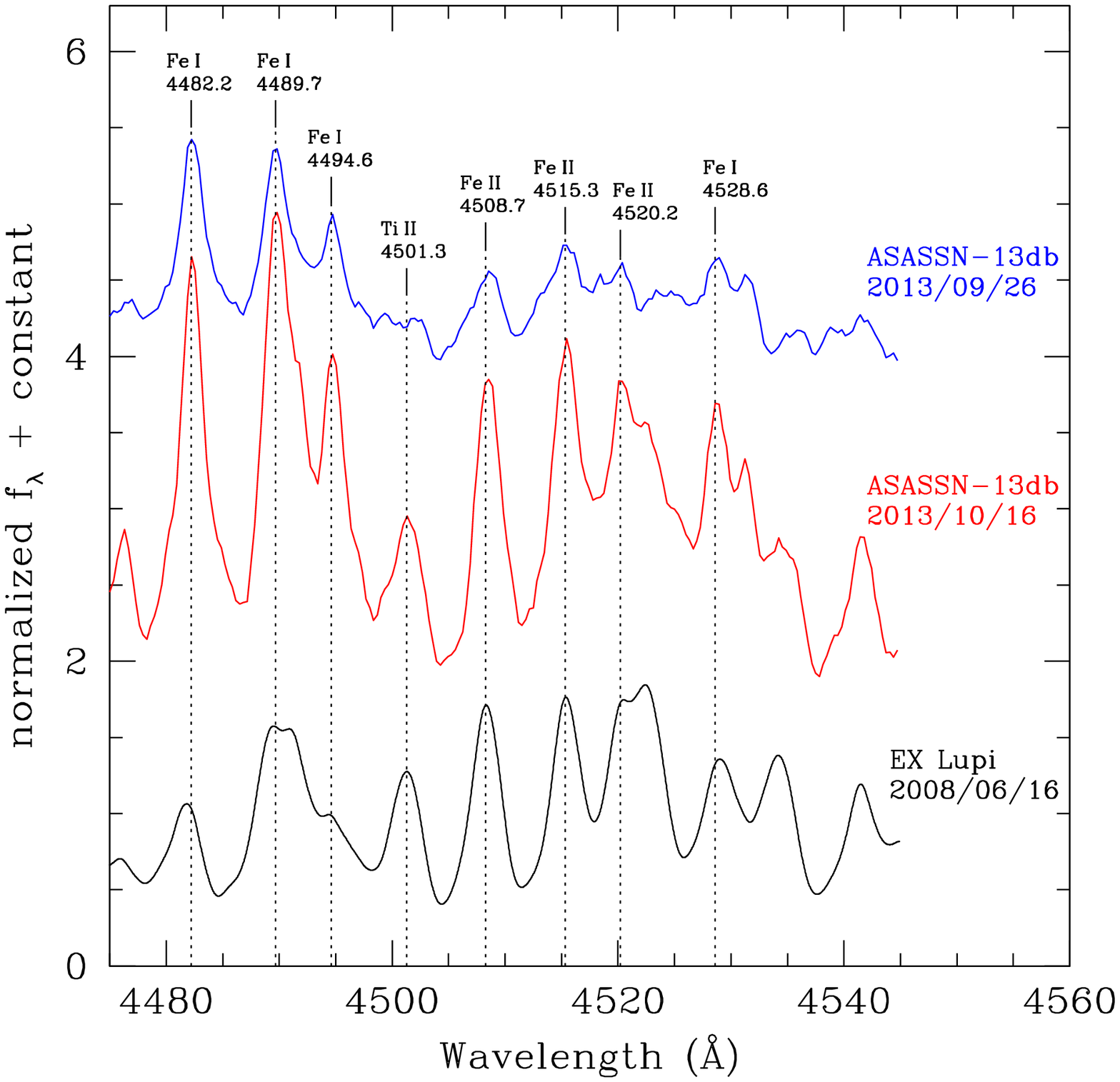}{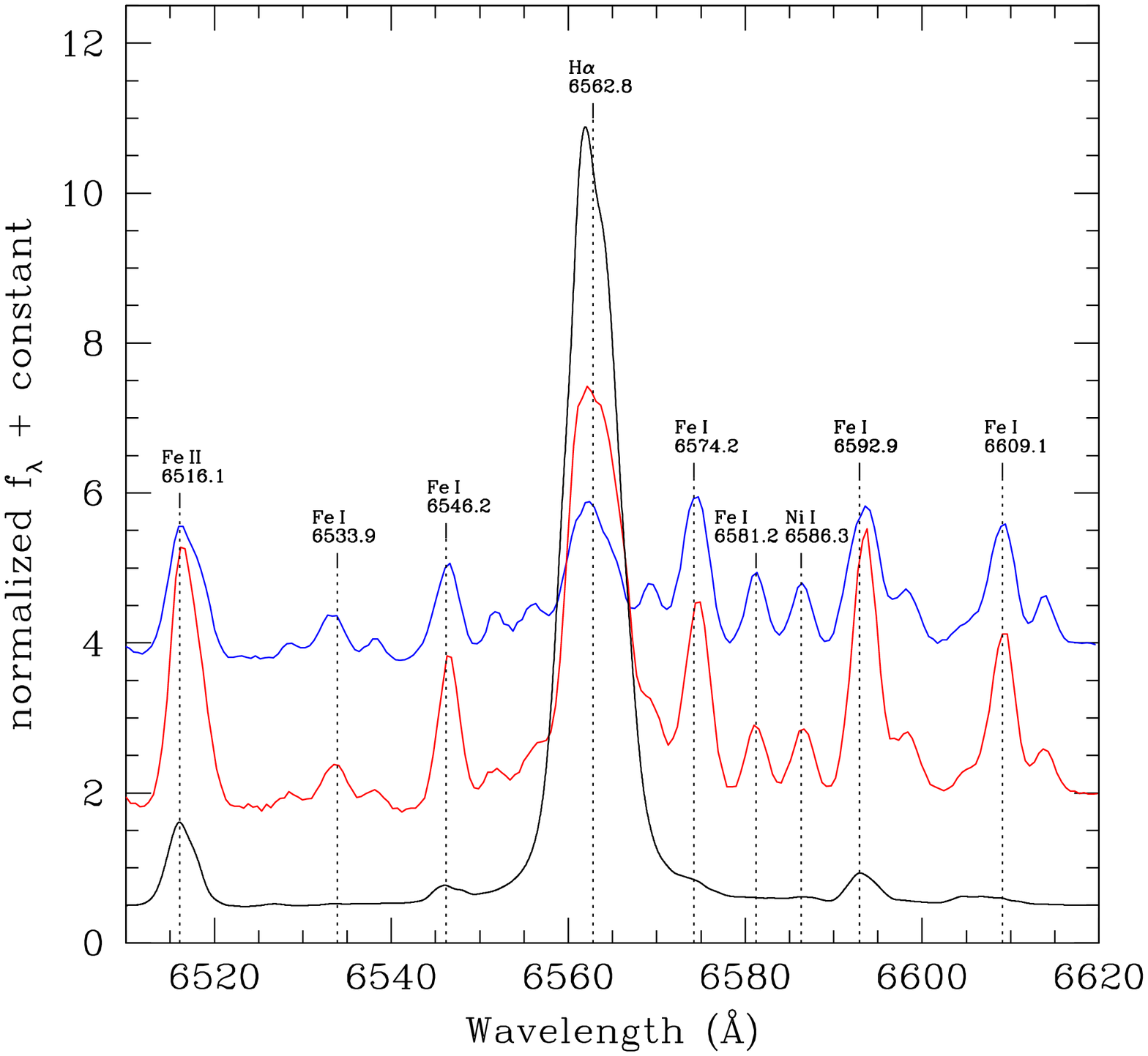} 
\caption{Continuum normalized spectra of {\name} compared to a spectrum of EX~Lupi during its latest and strongest eruption in June 2008 \citep{Sicilia2012}. We have convolved the high-resolution spectrum of EX~Lupi with a Gaussian to match the intermediate resolution of the Magellan/MagE spectra. All identified emission features of EX~Lupi are present in the {\name} spectra, but {\name} also shows many features not seen in EX~Lupi, some of which are identified in the figure  (e.g., the \ion{Fe}{1} lines at 6533.9~\AA~and 6581.2~\AA). We used a low-order polynomial to fit the minimal continuum between the emission lines and normalize the spectra.} 
\label{fig:speccomp}
\end{figure}

The emission line dominated spectra of ASASSN-13db are broadly similar to EXors in outburst, but with an unusually rich set of emission lines and little continuum. In Figure~\ref{fig:speccomp} we show two small sections ($4470-4560$~\AA~and $6510-6620$~\AA) of MagE spectra obtained at two different epochs separated by 20~days. For comparison, we also show a spectrum of EX~Lupi from \citet{Sicilia2012} obtained in June 2008 during its strongest observed outburst \citep[e.g.,][]{Kospal2008,Aspin2010}. We identify \ion{Fe}{1}, \ion{Ni}{1}, \ion{Fe}{2}, and \ion{Ti}{2} lines in emission, which are clearly present in the spectra of both ASASSN-13db and EX~Lupi. The metallic lines in the spectra of ASASSN-13db are resolved with $\rm FWHM \simeq 120-170$~km~s$^{-1}$, consistent with the broad components of the metallic emission lines in the spectra of EX~Lupi during its 2008 outburst. Our spectra of ASASSN-13db do not have enough resolution to show the narrow-line components observed in some emission lines during the EX~Lupi outburst \citep{Herbig2007,Sicilia2012}. {\name} does not show signs of large amplitude ($\gtrsim 5$~km~s$^{-1}$) radial velocity variations.

\section{Discussion}
\label{sec:disc}

Pre-outburst photometry and SED fitting indicate that {\starname} is likely a $L \simeq 0.1$~{\Lsun} TTS near the Orion star-forming region at a distance of $\sim420$ pc, consistent with its post-outburst spectrum. During the strong outburst from September 2013$-$January 2014, it increased in $V$-band flux by a factor of over 140 and maintained a level of emission $4-5.5$ magnitudes greater than in quiescence for much of this time. The magnitude of this outburst, its evolution, and its duration are consistent with those of the 2008 outburst of EX~Lupi, the prototype of the EXor variable class. In addition to this outburst, {\starname} also experienced a similar outburst in late-2010/early-2011. As EXors are known to experience recurring events every few years, this also suggests that {\starname} is an EXor.

Spectroscopy of {\starname} during outburst also shows similarities to the spectra of other EXors in the literature \citep[e.g.,][]{Sicilia2012, herbig08, lorenzetti12}. In particular, spectra of the source in outburst are dominated by a forest of narrow emission lines at all epochs, primarily \ion{Fe}{1} and other low-ionization lines. Comparison with spectra of EX~Lupi during its 2008 outburst from \citet{Sicilia2012} shows that the spectra of {\name} have many of the same emission features as EX~Lupi in 2008. However, the spectra of {\name} also show a number of lines not present in the EX~Lupi spectra, indicating possible differences in excitation temperature and/or abundances between the sources. As of late January 2014, the spectrum is that of an M5 TTS star.

The similarities between our observations of {\name} and EXors in the literature lead us to conclude that {\starname} likely underwent an EXor accretion event. The magnitude of the change in luminosity and the presence of many unique spectral features, however, are unusual for typical EXor outbursts, and more closely resemble the large 2008 outburst in EX~Lupi. If this was an EXor outburst, {\starname} would appear to be one of the lowest mass TTSs with such an event, and it seems to be experiencing large-scale events on a more rapid timescale than the typical EXor. 
\\
\acknowledgements
The authors thank D. Denisenko, A. Drake, L. Hartmann, S. Kenyon, S. Schmidt, and B. Skiff for useful comments and suggestions. We thank M. L. Edwards and the staff of the LBT Observatory for their support and assistance in obtaining the MODS spectra. We thank PI Neil Gehrels and the {\swift} ToO team for promptly approving and executing our observations. We thank LCOGT and its staff for their continued support of ASAS-SN.

Development of ASAS-SN has been supported by NSF grant AST-0908816 and the Center for Cosmology and AstroParticle Physics at the Ohio State University. JFB is supported by NSF grant PHY-1101216. JDS and JJA are supported by NSF grant AST-1108811.

This research has made use of the XRT Data Analysis Software (XRTDAS) developed under the responsibility of the ASI Science Data Center (ASDC), Italy. {\swift} at PSU is supported by NASA contract NAS5-00136.

This paper makes use of data from the AAVSO Photometric All Sky Survey, whose funding has been provided by the Robert Martin Ayers Sciences Fund.

The LBT is an international collaboration among institutions in the United States, Italy and Germany. LBT Corporation partners are: The University of Arizona on behalf of the Arizona university system; Istituto Nazionale di Astrofisica, Italy; LBT Beteiligungsgesellschaft, Germany, representing the Max-Planck Society, the Astrophysical Institute Potsdam, and Heidelberg University; the Ohio State University, and The Research Corporation, on behalf of The University of Notre Dame, University of Minnesota and University of Virginia.

This publication used data obtained with the MODS spectrographs built with funding from NSF grant AST-9987045 and the NSF Telescope System Instrumentation Program (TSIP), with additional funds from the Ohio Board of Regents and the Ohio State University Office of Research.


\end{document}